# A LITERATURE REVIEW ON INFORMATION SYSTEMS SUPPORTING THE PHYSICAL WELLBEING OF ELDERLY PEOPLE

Jose Teixeira
*Turku Center for Computer Science*
*Joukahaisenkatu 3-5, 20520 Turku, Finland*

Reima Suomi
*University of Turku*
*Rehtorinpellonkatu 3, 20500 Turku, Finland*

**ABSTRACT**

This paper reviews multi-disciplinary research on information systems supporting the physical wellbeing of the elderly population. By taking a systematic approach, it screens journals, conference proceedings and books on how computer-based systems are used for improving both the health and wellbeing of the most senior individuals. By using different Internet-databases indexing academic publications and a set of conceptual keywords the authors searched for and identified 62 major publications on the topic that were carefully reviewed.
Each publication item was classified according different category sets and the aggregated data was then analyzed from different socio-technological perspectives. Our findings suggest that research on the topic is very focused on diseases over health and wellbeing, since most of the studied information systems focused much more on the protective and curative medical procedures over other important dimensions such as prevention, education and health promotion. An overview on what and where the studied systems are used is presented and a new information systems research agenda is proposed.

## 1. INTRODUCTION

In a society aging at fast pace, computer-based technology must play an important role in supporting the elderly at coping with the natural but negative health impacts of growing older. After checking that existing literature reviews on technology supporting elderly people are scarce and outdate, the authors here perform an updated review on existing knowledge that points lenses at how computer-based technology is being employed to address the physical wellbeing of senior citizens.

The importance of literature reviews in the information systems field is highlighted by several authors . It is claimed that by seeing so few published literature review articles we are impeding the progress of the field. With this article we aim at facilitating theory development, identify areas where a plethora of research exists and uncover areas where research is needed within the scope of technology supporting the physical wellbeing of the most senior citizens. The authors, who had already previously reviewed a considerable amount of literature on the topic, decided to perform a more updated, systematic and reproducible literature review on the topic before embarking in new projects in the field.

Our research adopts the concept of physical wellbeing, as one of the several human health dimensions as proposed by the Salutogenesis and EUHPID public health models as described by Becker et al. (2010), Eriksson and Lindström (2008) and Bauer et al. (2003). In a triangle where the physical, mental and social wellbeing balance the individuals health, our research aggregates existing knowledge on computer-based technology supporting our older citizens in maintaining and recovering their physical conditions.

The degradation of the physical wellbeing affects each of us as soon as we are born, our society must accept it as a natural phenomenon, but it is a desirable goal in society to delay this natural phenomenon by employing preventive measures and, of course, recover everyone that typically after an accident or disease saw physical condition negatively affected.

Not many decades ago, the elderly were seen by the marketing departments of technology corporations as a small and uninteresting market. The older age-groups were a minority in the overall population; moreover they were associated as passive "clients" with low purchasing power (Phillips & Sternthal 1977), (Poulson 1997). Even worst, as narrated by at Zhou and Chen (1997), that reviewed the Canadian consumer magazine advertisement, business corporates completely ignored the elderly repeatedly over the past century when performing the marketing customer segmentation. Zhou and Chen (1997) also suggest that, when analyzing the impacts of different family roles in purchasing decisions, the grand-fathers and grand-mothers had less impact in the spending decisions that the fathers, mothers and teenagers.

Recently, elderly people have gained a better status as customers. The elderly are now seen as an interesting market by technology providers. Elderly are now a big part of the population, with considerable purchasing power and consuming more and more modern technology (Ijsselsteijn et al. 2007).

The paper unfolds as follows: First, and within the following section, the authors summary methodological and design issues that guided this literature review; secondly, we aggregate an analysis of relevant literature; and in final sections, we suggest key implications to the academic body of theoretical knowledge.

## 2. METHODOLOGY

### 2.1 Research goals and methodological base

This structured literature review aims to provide an aggregated vision of what is well known within the academia regarding computer-based technology supporting the physical wellbeing of elderly people..

This structured literature review took in consideration methodological issues provided by Webster and Watson (2002), Järvinen (2008) and von Brocke et al (2009). Maximum attainable transparency and rigor in documenting the literature review process, use of a systematic and future reproducible procedure were some of the base-pillars of this literature review. By taking von Brocke and Theresa (2011) and Lin (2011) as good examples of published literature reviews, it turned to be easier to perform the extraction, categorization and analysis of relevant research.

The literature review was performed between 16th December and the 27th of January 2012. The authors had previously reviewed literature on the topic, however by unstructured manners and not taking in account the scientific rigor of different sources. A lot of relevant literature was identified, but it matters to perform this structured literature review based on high-rigorous and peer-reviewed sources within the academia. Only basic and wide-available software tools such as a web-browser, spreadsheet software and a citation manager were used to search, filter, categorize and analyze the different literature.

### 2.1 Research design

The review process started by defining a research basis that guided how the literature was collected. The authors aimed at covering, by systematic manners, a broad range of literature by making use of general journal and conference databases available on the Internet. We started with the Emerald and EBSCO databases indexing mostly journal publications and later expanded our search to cover conferences and books by using the AIS electronic library and Google books system. More information on the used databases can be found in the following Table 1.

The authors, using advanced search mechanism, limited the search to peer-reviewed publications. Several published books and journal articles were discarded by not clearly demonstrating any evidence of conducting a peer-reviewed process. The search process was limited to research published in the English-language.

Table 1: Used general databases indexing journals, conference proceedings and books.

| Publications database | Description |
| --- | --- |
| Emerald insight | Index database of many academic publications by Emerald Group Publishing. Access granted via author's hosting university library services. |
| EBSCO Academic Search Premier | Index database of many academic publications by EBSCO Industries. Access granted via author's hosting university library services. |
| AIS Electronic library | Index of journals and conferences on Information Systems field. Freely available for members of the Association for Information Systems. Available at http://aisel.aisnet.org/. |
| Google books | Public domain books database provided by Google. Available at http://books.google.com. |

Several keywords were used to seek for literature addressing our investigation on computer-based systems supporting the health and wellbeing of the elderly population. Those search terms keywords are presented on Table 2. The decision of using several keywords benefited on the output of relevant research captured after the literature search process, however it increased the complexity during the search process: The high number of keywords forced the author to re-execute the same query several times just by changing keywords referring to the same terms.

Table 2: Keywords used

| Tokens for "Information System" | Tokens for "Elderly" |
| --- | --- |
| software | old/older |
| technology/technologies | elderly |
| system/systems | senior |
| computer/computers | aged |
|  | aging |
|  | senescent |
|  | third age |

After retrieving the results from the previous mentioned databases, the authors proceeded with an abstract analysis where most if the retrieved published items were discarded. Many items were discarded because they point to non-computer-based technology, therefore research on specific-purpose electronic and mechanical devices such as nurse calling systems or hearing loss electronics are out of scope of this literature review.

Many other retrieved items did not fit with our adopted conceptualization of physical wellbeing as a dimension of health. Considerable research exists on how elderly use, purchase or interact with computer-based systems but without referring any physical wellbeing implications.

After the filtering process the research basis was defined with two articles identified from the Emerald insight database, 58 different articles identified from the EBSCO Academic Search Premier database and two books retrieved from the Google books index of published books. No relevant publication were identified via the AIS electronic library. In Table 3 we present the hits gained from different search terms of the 62 identified research contributions.

Table 3. Total of selected literature review items as retrieved by keywords.

| 1$^{st}$ keyword IN title | 2$^{nd}$ keyword IN title | Total selected items |
| --- | --- | --- |
| technology(ies) | elderly | 24 |
| technology(ies) | older | 10 |
| technology(ies) | senior | 5 |
| technology(ies) | aged | 2 |
| system(s) | elderly | 12 |
| system(s) | older | 3 |
| system(s) | senior | 3 |
| system(s) | aged | 1 |
| software | elderly | 2 |
| TOGETHER |  | 62 |

With a selection of 62 peer-reviewed publications on the topic, we proceeded then to a careful reading, analysis and categorization of research as described in the following section.

## 3. ANALYSIS AND CATEGORIZATION OF RELEVANT RESEARCH

After a fast reading of the initial set of 62 papers, the authors decided to drop 12 papers due to its inadequate fit with our research purposes. When evolving from abstract and partial paper reading to a more careful reading, the authors were forced to drop Ko (2003), National Institute of Building Sciences (1996), Dunkle & Haug (1984), Shellenbarge (2002), Shellenbarger(1999), Rabner (1999) and Zen (2007) either because those publications were not peer-reviewed as initially appeared or because its perceived rigor was inadequate to the target of the research being conducted. Moreover, the publication Lubinski & Higginbotham (1998) Figueiro et al. (2008), Goins et al. (2010), Bailey et al. (2011) and Lesner & Klingler (2011) were left out because they were not dealing with computer-based technology as initially appeared, but instead with specific electronics, mechanical systems or lightning technology. After reading all papers we were left with a set of 50 for a careful analysis and categorization.

From the set of resulting papers, and without looking at its content, some interesting remarks should be pointed out: First, by analyzing the authors affiliations we can observe that most research on the topic was produced in North America, Northern Europe, Spain, Hong-Kong, Japan and Korea. From a set 50 papers, 14 publications came solely from the USA, 8 from Sweden and 7 from UK; with this three countries accounting for circa 60% of identified research on the field of Information Systems supporting the elderly; Secondly, there is a lack of research collaboration between researchers from different countries, only five of the publications withing the set where published with authors from different countries. European union research funding initiatives where behind most of academics collaboration on the field, but we can suggest that countries are duplicating research efforts over collaborating. Finally, the research identified on this literature is clearly mulls-disciplinary as it was published in a large and diverse set of journals. No journal contributed with more than two articles for the identified research basis set.

Each of the identified papers was careful read. Then, the authors extracted each paper research questions, together with both theoretical and practical implications to a different medium. At a later stage, the papers were then grouped regarding their methodological approaches, and their theoretical or empirical relevance. A last categorization was made based on the dimensions of technology, clinical use and relation to patient. It is important to notice that authors did not start with a rigid set of categories for classifying the identified publication items, those categories been keeping adjusted over the literature review process.

In the following section, we present some relevant socio-technological patterns and characteristics for the reviewed literature, for a information-system research point of view.

## 4. RESULTS AND FUTURE WORK

Most of the found literature was conducted by qualitative and descriptive manners. Plenty of theories and frameworks are ready to be tested and challenged by more quantitative approaches, those last approaches seem to be lacking on the studied area. An important issue emerged when the authors classified the reviewed literature items according the Eriksson and Lindström (2008) stages interpreting the *salutogenic* public-health theory by Antonovsky (1987). As seen in Figure 1 the reviewed literature targets disease-care over health-care, since most of the studied information systems focused much more on protective and curative medical procedures over other important dimensions such as prevention, education and health promotion. The authors strongly believe that there is plenty of space for academics and practitioners to research and develop preventive and well-being information systems over the most reactive and traditional health informatics approach.

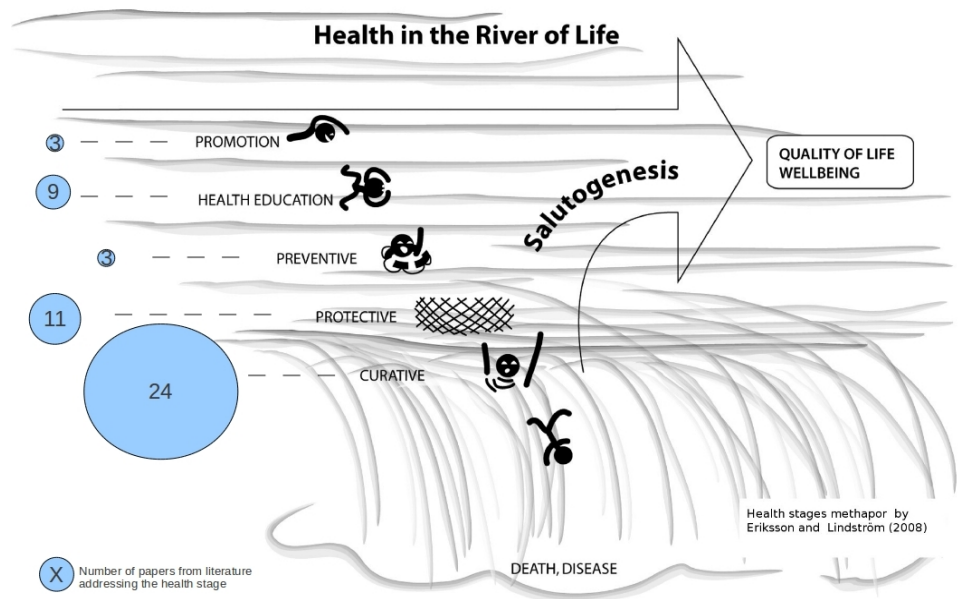

Figure 1: Literature among the different *salutogenic* public-health stages

While reviewing literature the authors identified and clustered, from a technological perspective, different information systems supporting the elderly. Only after reviewing all the selected literature, the authors decided for grouping the published research in a taxonomy with seven technology-types; As seen in Table 4 the literature was grouped as following: a first set includes database, decision support systems (DSS) and data analytics systems widely infused in the modern enterprise; then tele-medicine technologies aimed at providing clinical health-care at by distance; assistive technology aimed at supporting people with disabilities at their own home environment; Information portals for disseminating wellbeing knowledge; Alarms and other systems triggering emergency care, Robots performing specific health-related operation; and finally, a last category, for more diverging and disruptive technologies, i.e. electrocardiogram analysis , used as well to classify published literature covering many types of technology.

Table 4: Technology clustering of IS supporting the elderly

| Year | Database, DSS and analytics | Tele-medicine | Assistive technology | Information portals | Alarm and emergency systems | Robotics | Others & many |
|---|---|---|---|---|---|---|---|
| 1985 | | | | | | | Gilly & Zeithaml |
| 1986 | | | | | | | |
| 1987 | | | | | | | U.S. Congress, Office of Technology Assessment |
| ... | | | | | | | |
| 1995 | Love & Lindquist | | | | | | |
| ... | | | | | | | |
| 1998 | | | | | | | Eby & Kostyniuk |
| 1999 | | Severs | | | Other | | |
| 2000 | | | | Molloy et al. | | | |
| 2001 | | | | | | | |
| 2002 | Logue | | | | | | Dozet et al. Lubinskic & Higginbotham |
| 2003 | Cortés et al. | | Lilja et al. Bradley et al. Poppen Hoenig et al. | | | | |
|      |                    | Bodoff         |                 |                 |              |          |                      |
| 2004 |                    |                |                 |                 | Hyysalo      |          | Östlund              |
| 2005 |                    |                |                 |                 |              |          | Merat et al.         |
| 2006 | Downs et al.       |                |                 |                 |              |          | Dethlefs & Martin    |
| 2007 | Hedström           | Bertera et al. |                 |                 |              |          | Mant et al.          |
|      |                    |                |                 |                 |              |          | Sokoler & Svensson   |
| 2008 | Dorr et al.        | Botsis et al.  |                 | Torp et al.     |              |          |                      |
|      |                    |                |                 | Vimarlund.      |              |          |                      |
|      |                    |                |                 | Tse et al.      |              |          |                      |
| 2009 | Botella et al.     | Londei et al.  |                 |                 | Leung        | Yamauchi |                      |
|      | Kishimoto et al.   | Lai et al.     |                 |                 |              |          |                      |
|      | Krienert et al.    |                |                 |                 |              |          |                      |
|      | Rajasekaran et al. |                |                 |                 |              |          |                      |
| 2010 | Fejes & Nicoll     |                | Harrefors et al.| Fernando et al. |              |          | Huber                |
|      | Khan et al.        |                |                 | Iliffe et al.   |              |          | Korpinen & Pääkkönen |
|      | Kasteren et al.    |                |                 | Goodall et al.  |              |          |                      |
|      | Kang et al.        |                |                 |                 |              |          |                      |
| 2011 | Fossum et al.      |                | Zwijsen et al.  | Hanson et al.   |              |          |                      |
|      | Ishigaki et al.    |                | Miskelly        |                 |              |          |                      |
|      | Etchemendy et al.  |                |                 |                 |              |          |                      |
|      | Lim et al.         |                |                 |                 |              |          |                      |
|      | Carlson et al.     |                |                 |                 |              |          |                      |

From the reviewed literature we suggest that most computer-based technology, seeking benefits for the elderly population, seems designed to be used by health-care professionals and not by elderly themselves. In many of the database, DSS, analytics, monitoring and assistive technology systems; there is an extremely low patient empowerment: in most cases patients are merely input data for computer-based systems, health-care professionals and police-makers; very few of the reviewed information systems seems to be designed for benefiting the elderly individuals directly. With this low empowerment of senior citizens over other actors in the national health system, the realization of technological benefits will be on the hands of health-care professional and public-health officers. The elderly, seem to not have a choice in taking their own responsibility on using ,by themselves, information systems that could improve their wellbeing.

Most of the information systems covered by the studied literature are deployed at physical installations of hospitals, day-care centers and public-health bodies. On the other hand, identified research on the field of assitive technology brings the systems to elderly homes, however with strict and persistent connections with central systems monitoring elderlies activity. The is a trend for integration on many of those different systems using Internet as a hub, a current and promising field of research for both academic researchers and practitioners.

For future work, an reflecting relevant literature retrieved by non-systematic manners, the authors plan to extend the research basis to include relevant papers cited withing the careful reviewed items so far. Moreover, the authors plan to extend the research basis to include more journals and conferences not indexed by the already used databases indexing academic publications.

## 5. CONCLUSION

Multi-disciplinary research on the investigated topic made spectacular developments in the last two decades. The phenomenon seams to have gained considerable attention from researchers between 2007 and 2011; However, its evident that we are still at the beginnings of using information systems for helping the most elderly population.

More empirical observations on the phenomenon should be reported by researchers together with and more rigorous research testing existing body of knowledge is needed. Developers of such technology have plenty of unexplored areas such as patient-to-patient wellbeing, patient empowerment, user-centric assistive technology, chronic-care expert systems, specific-purpose robotics, etc. With a society aging at fast pace, we urge researcher and practitioners to embrace this multi-disciplinary are of research with many research gaps to be filled.